# Q-Graph: Preserving Query Locality in Multi-Query Graph Processing

Christian Mayer, Ruben Mayer, Jonas Grunert, Kurt Rothermel, and Muhammad Adnan Tariq
Institute of Parallel and Distributed Systems, University of Stuttgart, Germany
{firstname.lastname}@ipvs.uni-stuttgart.de

## ABSTRACT

Arising *user-centric* graph applications such as route planning and personalized social network analysis have initiated a shift of paradigms in modern graph processing systems towards *multi-query analysis*, i.e., processing multiple graph queries in parallel on a shared graph. These applications generate a dynamic number of *localized* queries around query hotspots such as popular urban areas. However, existing graph processing systems are not yet tailored towards these properties: The employed methods for graph partitioning and synchronization management disregard query locality and dynamism which leads to high query latency. To this end, we propose the system *Q-Graph* for multi-query graph analysis that considers query locality on three levels. (i) The query-aware graph partitioning algorithm *Q-cut* maximizes query locality to reduce communication overhead. (ii) The method for synchronization management, called *hybrid barrier synchronization*, allows for full exploitation of local queries spanning only a subset of partitions. (iii) Both methods adapt at runtime to changing query workloads in order to maintain and exploit locality. Our experiments show that Q-cut reduces average query latency by up to 57 percent compared to static query-agnostic partitioning algorithms.

## 1 INTRODUCTION

Distributed graph processing systems such as Pregel [22], PowerGraph [12], and PowerLyra [5] have emerged as the de facto standard for batch graph processing tasks due to their superior performance of data analytics on graph-structured data. To parallelize graph execution on $k$ workers, these systems divide the graph into $k$ partitions such that the number of vertices or edges are balanced and localized, i.e., neighboring vertices and edges preferably reside on the same partition. However, novel graph applications have given rise to a shift of paradigms towards interactive graph queries on a shared graph structure [9, 44]. These applications share three properties to which existing graph partitioning algorithms are not tailored. First, the scope of the queries is localized, variable, and overlapping (i.e., the query workload is *clustered* with computational hotspots in the graph). Second, multiple iterative and long-running graph queries run in parallel on a shared graph. Third, there are variations of query workload and hotspots. We denote these applications as **c**oncurrent **g**raph query **a**nalytics (**CGA**). In this paper, we address the question: how to tailor graph partitioning and management to CGA applications in order to reduce query latency? To answer this, let us first examine three typical CGA applications.

**Application 1:** Today, many users request mapping services such as {Google, Apple, OpenStreet}-Maps to perform route planning computations that can be modeled as shortest path queries with start and end vertices on a huge road network [46]. These queries are inherently localized (*what is the shortest path from home to my doctor?*) – more than 50% of mobile search queries have local intent [3]. This leads to computational hotspots in certain graph areas (e.g. urban centers) that are subject to short- and long-term changes (e.g. festivals and growing cities), as well as regular fluctuations of query workload with the time of the day or the day of the week. To yield high customer satisfaction, these services must serve multiple graph queries in parallel, with minimal query latency.

**Application 2:** Digitized social networks allow users to measure the impact of their online activity [11]. In general, a user accesses only a personal social network (i.e., *social circle*) – to maintain privacy (e.g. Facebook restricts visibility of posts), to protect integrity and trust, or to personalize usage [20]. The social circles overlap because of the high clustering coefficient of social networks [40]. Therefore, multiple graph queries, that are localized and overlapping, access shared graph data in social network analysis such as influence propagation [2, 18], information dissemination [4], community detection [31], and friendship recommendation [13]. Moreover, there are computational hotspots around hubs [27] that change over time (e.g. changing popularity of a star).

**Application 3:** Large-scale knowledge graphs store structured knowledge and enable retrieval of information [17]. For instance, smart cars use SPARQL [29] queries on knowledge graphs to match control rules [10]. These types of graph queries access only a small portion of the graph, but there are multiple parallel queries (e.g. from multiple smart cars) leading to computational hotspots around content with dynamic popularity.

We identified three challenges for CGA applications. (i) *Locality:* how to ensure locality of query execution? To parallelize query execution, graph systems partition the graph across $k$ workers. Hence, *queries are distributed across workers* which slows down query execution due to network communication and synchronization overhead between workers. A major challenge is to develop a partitioning algorithm that considers query workload to improve locality, scalability, and latency of the graph system. (ii) *Multi-query:* how to manage the execution of multiple parallel queries on a shared graph? Single-query systems use a global barrier synchronization to execute the graph query in an iterative manner. After each iteration, all vertices have finished execution before proceeding with the next iteration [22]. However, multi-query systems require carefully designed synchronization mechanisms to decouple query execution and to fully exploit the query locality on the partitioned graph. (iii) *Adaptivity:* how to adapt to dynamic query hotspots and query workloads? A key challenge is to repartition the graph at runtime to address changing query hotspots.

| | Pregel, Kineograph [6, 22] | PowerLyra, GraphX [5, 43] | GrapH, Mizan [19, 26] | Weaver, Seraph [9, 44] | Q-Graph |
|---|---|---|---|---|---|
| Locality | ✗ | ✓ | ✓ | ✗ | ✓ |
| Multi-query | ✗ | ✗ | ✗ | ✓ | ✓ |
| Adaptivity | ✗ | ✗ | ✓ | ✗ | ✓ |

Table 1: Research gap: multi-query graph processing with adaptive partitioning maximizing query locality.

In Table 1, we categorize related work with respect to these challenges. We identified a research gap with respect to a *multi-query graph system with adaptive partitioning*. Existing graph systems such as Pregel [22], Kineograph [6], PowerLyra [5], and GraphX [43] do not support execution of multiple queries in parallel and are not adaptive to changing graph workload. Although GrapH [27] and Mizan [19] are adaptive, they do not support parallel queries. Existing multi-query systems such as Seraph [44] and Weaver [9] do not partition the graph in a locality-preserving manner.

This paper introduces the open-source system Q-Graph [24] for CGA applications that fills this research gap. Q-Graph uses a centralized controller to maintain global knowledge about the queries running on each worker to perform *high-level, adaptive, query-aware partitioning*. We show that query-awareness of partitioning and synchronization speeds up CGA applications compared to query-agnostic static partitioning algorithms—as a result of improved query locality and workload balancing. In particular, this paper provides the following contributions:

- A high-level query-aware partitioning algorithm *Q-cut* that partitions the graph based on a history of queries and reaches high query locality of up to 80%. Query-aware partitioning is fast because it operates on a small number of queries rather than a large number of vertices (cf. Section 3.2).
- A hybrid-barrier synchronization model with local *and* global barriers. Local barriers reduce query latency for localized queries due to minimal synchronization overhead, while global barriers enable graph management for repartitioning and adaptivity (cf. Section 3.3).
- An adaptation method that optimizes the graph partitioning at runtime to dynamic query hotspots using the centralized global knowledge regarding query locality (cf. Section 3.4).
- A thorough evaluation of our system Q-Graph showing reduced query latency by up to 57% compared to state-of-the-art partitioning and by up to 47% compared to state-of-the-art barrier synchronization (cf. Section 4).

## 2 PROBLEM DESCRIPTION

In this section, we introduce the background, notations, and the partitioning problem.

**Background and Notations:** The graph structure is given by $G = (V, E)$ with the set of vertices $V$ and the set of directed edges $E \in V \times V$. Each vertex $v$ maintains vertex data $D_v$ (edge data is stored within the vertex data). We follow the predominant *vertex-centric programming model* where each vertex iteratively recomputes its own vertex data based on messages from neighboring vertices. Vertex $v$ exchanges data with a neighboring vertex $v'$ by sending a message $m_{v \to v'}$. The application programmer specifies the vertex function $f(D_v, m_{* \to v})$ for vertex $v$ and the set of incoming messages $m_{* \to v}$. We define a query $q$ as a tuple $(f, V_{sub})$ of a vertex function $f$ and an initial subset of active vertices $V_{sub} \subseteq V$. An example is the problem of finding the shortest path between the start vertex $v_0$ and the sink vertex $v_{end}$. The initial subset of vertices contains only the start vertex $v_0$, i.e., $V_{sub} = \{v_0\}$. The query function for a given vertex $v$ iteratively recalculates the shortest distance from the start vertex $v_0$ to vertex $v$ by updating the distance based on the distance of neighboring vertices (cf. Single-Source-Shortest-path (SSSP) [44]).

Multi-query graph systems support two types of requests: read-only graph analytics queries and write-enabled graph updates [8, 9, 44]. Graph analytics queries (i.e., *queries*) can read the complete vertex data but write only on separate query-specific vertex data to prevent a write conflict between any pair of queries. We focus on the efficient parallel execution of multiple graph analytics queries.

To enable a consistent view on the vertex data, batch graph processing systems execute graph vertices using the synchronous and iterative bulk synchronous processing (BSP) model [22, 27, 35] consisting of three phases: i) computation, ii) communication, and iii) barrier synchronization. In the computation phase, active vertices calculate the value of the updated vertex data based on past messages. In the communication phase, active vertices asynchronously send messages to neighboring vertices that may reside on different workers. In the synchronization phase, the system waits for all vertices to finish phases i) and ii) (*barrier synchronization*). A vertex $v$ is considered as active in iteration $i$ if any other vertex has sent a message to vertex $v$ in iteration $i - 1$. These three phases, denoted as one *iteration*, are repeated until no active vertex remains.

**Dynamic Partitioning Problem:** The graph structure provides information about data dependencies: graph vertices exchange messages only with neighboring vertices. Hence, sophisticated partitioning algorithms exploit the graph structure to minimize the number of messages that are sent across partition boundaries. Higher data locality improves scalability, latency, and communication overhead [27]. However, when executing multiple queries in parallel, the partitioning problem formulation has to be extended.

Given the graph $G$, a set of queries $\mathbb{Q} = \{q_1, ..., q_p\}$, and a pool of workers $W = \{w_1, ..., w_k\}$. Roughly speaking, the goal is to find an assignment of vertices to workers at each point in time such that the average query latency is minimal. The average query latency is influenced by the partitioning of the active query vertices: High locality of query vertices reduces the latency overhead for network communication. The dynamic partitioning is given by an assignment function $\mathcal{A}$ that assigns vertices to workers at different points in time, i.e, $\mathcal{A} : V \times T \rightarrow W$ for the set of vertices $V$, a time interval $T = [T_{min}, T_{max}] \subset \mathbb{R}$, and the set of workers $W$. Let a function $X : \mathbb{Q} \times T \rightarrow \mathcal{P}(V)$ track the set of active vertices for a query and a point in time. The function $X$ incorporates the impact of partitioning decisions on the query latency – higher locality reduces message delay which leads to earlier activation times of vertices. The average query latency is given by the (averaged) difference between the last and the first time instance in which a query has at least one active vertex, i.e., $\frac{1}{|\mathbb{Q}|} \sum_{q \in \mathbb{Q}} max\{t \in T | X(q,t) \neq \emptyset\} - min\{t \in T | X(q,t) \neq \emptyset\}$.

To measure *partitioning quality* for CGA applications, we define the *global query scope* $GS(q) \subseteq V$ as the set of all vertices that are activated during execution of query $q$ within a time interval of $\mu$ seconds. The value of $\mu$ captures the sensitivity to different activation patterns and is discussed in Section 3.4. Likewise, we define the *local query scope* $LS(q, w) \subseteq GS(q)$ as a subset of vertices in the global scope of query $q$ that are assigned to worker $w$. If $LS(q, w) = \emptyset$, query $q$ has not activated any vertex on worker $w$. Furthermore, if $LS(q, w) = GS(q)$, query $q$ is completely local on worker $w$. Using these definitions, we define the *query-cut* metric measuring the quality of a given partitioning, i.e., the *locality of query execution*. More formally, query-cut is the number of non-empty local query scopes $\sum_{q \in \mathbb{Q}} |\{w \in W | LS(q, w) \neq \emptyset\}|$.

We solve the problem of *balanced k-way query-aware partitioning* that minimizes the query-cut as defined previously while ensuring balanced partitions with respect to the number of active vertices (cf. Section A.1). A smaller query-cut increases query locality and decreases query latency due to the reduced communication and synchronization overhead. Intuitively, executing a graph query on a single worker is fast (cf. Section 4.2) – a worker iterates over local vertices and executes the vertex function without waiting for remote workers, without overhead for serializing and deserializing messages, without passing the multi-layered TCP/IP stack through the operating system, and without the delay of transferring data over the network.

**Research Gap:** Existing partitioning methods, such as *balanced k-way partitioning* for $k$ workers [36], are agnostic to the dynamic query workload. Graph partitioning is either based on edge-cut [36, 38] or vertex-cut [12, 27], i.e., minimizes the number of adjacent vertices or edges that are assigned to different workers. However, for CGA applications, data dependencies are not necessarily defined by the graph structure but, to a large extent, by the *localization of the graph queries*. In Figure 1, we give an example on the neighborhood graph of New York districts with two localized queries $q_1$ and $q_2$ running on the same graph. A minimal 2-way query-aware partitioning would partition the graph such that no query is divided into multiple parts. Hence, any cut that separates queries $q_1$ and $q_2$ is minimal and leads to zero traffic between the partitions at a certain time instance. However, the *query-agnostic* edge-cut partitioning algorithm would prefer *cut 1* with edge-cut size six over *cut 2* with edge-cut size eight – and invest scarce computational resources to calculate this cut. Therefore, query-agnostic partitioning algorithms traverse larger search spaces without producing better partitioning quality for CGA applications. Even worse, a query-agnostic edge-cut algorithm prefers suboptimal *cut 3* with edge-cut size two over *cuts 1 and 2* leading to a more expensive distributed query execution.

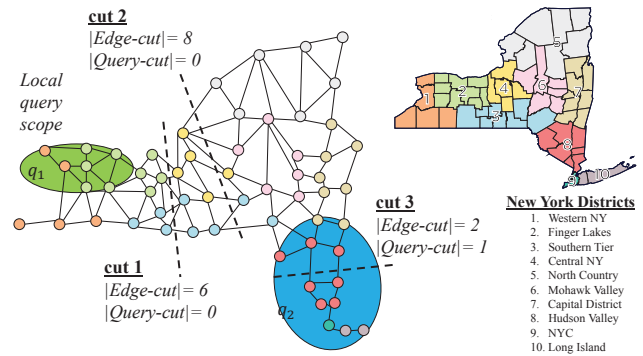

**Figure 1: Query-agnostic partitioning optimizes edge-cut and query-aware partitioning optimizes query-cut [39].**

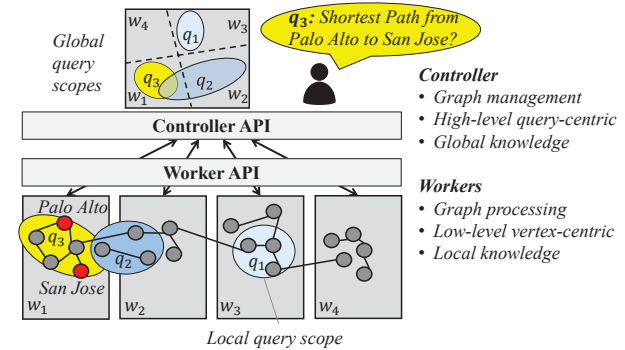

**Figure 2: System Architecture.**

## 3 Q-GRAPH SYSTEM

In this section, we provide an overview of our distributed graph system Q-Graph for parallel graph query processing and describe three optimizations: query-aware partitioning, hybrid barrier synchronization, and query-aware adaptivity.

### 3.1 System Overview

Q-Graph consists of a two-layered architecture: the worker and the controller layer (cf. Figure 2). The workers perform distributed graph query processing, i.e., they execute the vertex functions on the active vertices and handle message exchanges between neighboring vertices residing on different workers. The centralized controller manages execution of the graph queries on the distributed graph using a scalable representation of global knowledge about the graph workload. It dynamically adapts the partitioning at runtime and ensures efficient (barrier) synchronization of the graph queries. We developed a general-purpose API for both the worker and the controller layer (cf. Table 2). The controller provides a front-end for users to access graph processing resources with the request $scheduleQuery(q)$ (e.g. the user schedules query $q_3$ in Figure 2). The controller handles execution of this query by forwarding the scheduling request to the workers (i.e., calling $executeQuery(q)$). The next three sections describe the query-aware algorithm Q-cut (cf. Section 3.2), the hybrid barrier synchronization (cf. Section 3.3), and the adaptivity method (cf. Section 3.4).

| Command | Description |
| --- | --- |
| **Controller API** | |
| $stats(q, |LS(q, w)|, I_w, w)$ | Worker $w$ updates controller with statistics about $q$'s local query scope and intersection $I_w$. |
| $barrierSynch(q, w)$ | Worker $w$ indicates termination of current iteration of query $q$. |
| $scheduleQuery(q)$ | User schedules query $q$. |
| **Worker API** | |
| $move(LS(q, w), w, w')$ | Controller requests worker $w$ to move $q$'s local scope to $w'$. |
| $barrierReady(q)$ | Controller releases worker waiting for $q$ barrier. |
| $executeQuery(q)$ | Controller requests worker to start query $q$. |

**Table 2: Q-Graph API.**

### 3.2 Q-cut: Centralized Query-aware Partitioning

Q-Graph pushes partitioning decisions to the controller to benefit from the global knowledge about the query scopes and workload. The problem is that maintaining low-level global knowledge about the complete graph (vertices and edges) is not scalable. To this end, one of our central ideas is to use a more scalable representation of global knowledge by focusing on high-level query scopes rather than low-level vertex information. As the number of queries is much smaller than the number of vertices, the controller can utilize powerful partitioning algorithms on small data to improve query locality and workload balancing.

*3.2.1 Partitioning Strategy.* Our strategy for query-aware partitioning involves three steps (cf. Figure 3) organized as a MAPE loop: monitor-analyze-plan-execute. First, workers update the controller with query statistics after every iteration: for each active query, each worker sends the size of its local scope and the size of the *overlap* with other local query scopes. Hereby workers transform their low-level knowledge (i.e., *which* vertices each query executes) into high-level knowledge (i.e., *how many* vertices each query executes). In the example, there are two workers $w_1$ and $w_2$ and three queries $q_1, q_2, q_3$ on the New York districts graph from Section 2. Queries $q_1$ and $q_3$ span both workers leading to communication in each iteration of the query. Queries $q_2$ and $q_3$ overlap on worker $w_2$. Second, the controller performs query-aware partitioning (Q-cut) on the high-level representation with reduced problem size to find high-quality query-cuts. Third, the controller transforms the resulting query-cut back to the low-level representation into a proper graph partitioning by sending *move* requests to workers. In the example, local query scope $LS(q_1, w_2)$ moves to worker $w_1$ and local query scope $LS(q_3, w_1)$ moves to worker $w_2$. The new partitioning contains no queries spanning multiple workers and, therefore, has perfect query locality.

*3.2.2 Q-cut Algorithm.* The goal of query-aware partitioning is to maximize query locality. To this end, we define a cost function $c : \mathbb{S} \to \mathbb{R}$ in the space of potential solutions $\mathbb{S}$ of valid Q-cuts. We set the cost function $c_s$ for solution $s$ to the following query-cut metric:

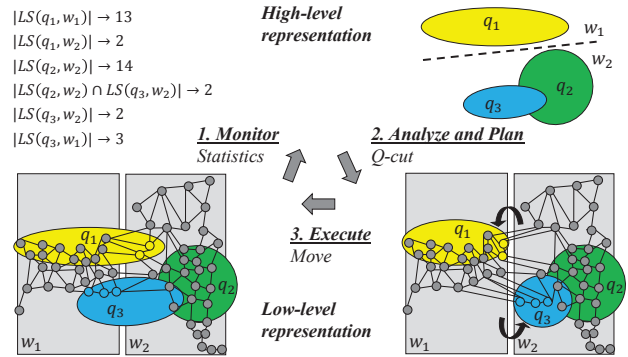

**Figure 3: The Q-cut algorithm operates on a high-level query representation to improve low-level graph partitioning.**

we sum over all queries $q_i \in \mathbb{Q}$ the number of vertices that are not assigned to the worker with the largest query scope for query $q_i$, i.e., $\sum_{q_i \in \mathbb{Q}} \sum_{w \in W, w \neq argmax_{w' \in W} |LS(q_i, w')|} |LS(q_i, w)|$. For example if two workers execute two queries completely independently, the costs would be zero.

The algorithm should (a) retrieve low-cost solutions effectively when the controller has enough time (i.e., the problem size is small enough to be solvable within the time limit), (b) provide the best found solution when interrupted, and (c) find low-cost solutions for diverse query workloads (i.e., does not overfit to specific problems).

---

**Algorithm 1** Iterated local search algorithm for Q-cut partitioning.

1: state $\hat{s} \leftarrow$ INITIALSOLUTION()
2: **while** not TERMINATED() **do**
3:    $s \leftarrow$ PERTURBATION($\hat{s}$)
4:    $s \leftarrow$ LOCALSEARCH($s$)
5:    **if** $c_s < c_{\hat{s}}$ **then**
6:       $\hat{s} \leftarrow s$

---

A well-established algorithmic framework for optimization problems meeting these requirements is *iterated local search* (ILS), a meta-heuristic that generates a sequence of solutions – each building on top of the previous solution [21]. For ILS, we must define the cost function $c_s$ for each solution $s$ and provide four subroutines: (i) the initial solution, (ii) a local search heuristic to find a local minimum given an initial solution, (iii) a perturbation method to overcome local minima, and (iv) a termination criterion (cf. Algorithm 1). After generating an initial solution, we iteratively perturb the current solution to avoid getting stuck in local minima and perform the local search method until the next local minimum is found. We store the solution with minimal costs in the variable $\hat{s}$. Selecting suitable subroutines is crucial for effectiveness and efficiency [21]. In Appendix A, we describe the four subroutines in detail. Note that all solution states have balanced workload (cf. Algorithm 2).

### 3.3 Hybrid Barrier Synchronization for Multi-Query Graph Processing

To perform synchronous query execution of query $q$, each worker $w$ informs the controller via the API call $barrierSynch(q, w)$ that all active query vertices have terminated in this iteration. After this, the worker waits for the *barrierReady(q)* message from the controller before starting with the next iteration. As soon as the controller has received all barrierSynch-messages, it sends the *barrierReady(q)* message to all waiting workers.

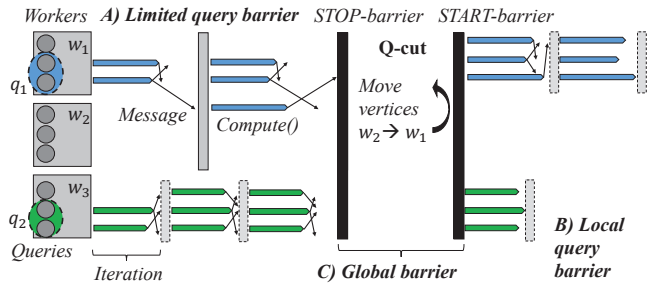

**Figure 4: The hybrid barrier synchronization integrates: A) limited and B) local query barriers, and C) global barriers.**

The state-of-the-art synchronization model [44] introduces an independent global barrier for each query. This leads to two problems: (1) redundant *global* barriers cause communication overhead, (2) local queries, which could be executed on a single worker, face unnecessary global synchronization.

Similar to Xue et al. [44], we decouple execution of multiple queries using an independent barrier for each query (denoted as *query barrier*). The alternative, i.e., a barrier that is shared by all queries, would result in a straggler problem because all queries have to wait for the slowest query *after each iteration*. However, to avoid unnecessary global synchronization for queries that run on a subset of partitions, we introduce *local query barriers* and *limited query barriers* which synchronize only between workers that are currently involved in the query execution.

We developed a three-phased synchronization mechanism denoted as *hybrid barrier synchronization*. 1) We decouple query barriers to mitigate the straggler problem and enable independent query execution. 2) We introduce the *limited query barrier* to prevent synchronization between workers that do not execute the same query. The most extreme limited barrier is the *local query barrier* that allows communication-free execution as long as queries remain local, i.e., no distant vertices get activated via message passing. 3) In regular intervals, we initiate a global barrier that is shared across all workers. The global barrier consists of a *STOP-barrier* that halts the whole system and enables global optimization of the partitioning and a *START-barrier* that resumes normal query execution on the optimized system after all optimizations are implemented.

In Figure 4, we exemplify the execution of two queries $q_1$ and $q_2$ on a graph that is partitioned across three workers $w_1, w_2, w_3$. We indicate the computation time of a single vertex as a horizontal bar in the respective query color. Initially, both queries are local. But after the first iteration, query $q_1$ activates a neighboring vertex on worker $w_2$ which initiates a limited barrier between workers $w_1$ and $w_2$ leading to the exchange of barrier messages between the workers and the controller (not shown). After three iterations of query $q_2$, the controller decides to initiate a global barrier to perform repartitioning. As a result, query $q_1$ switches to local execution mode on worker $w_1$.

### 3.4 Adapting to Dynamic Query Workload

CGA graph systems are subject to changing query workload patterns and query hotspots are not known in advance. To enable adaptivity of Q-Graph, each worker $w$ keeps the controller updated about the sizes of the local query scopes and their intersections by regularly sending *stats* messages. The controller combines the size of the local query scopes into the global query scope size and initiates repartitioning decisions by calling $move(LS(q, w), w, w')$ to move the local scope of query $q$ from worker $w$ to $w'$ (i.e., fuse both local scopes of query $q$ on worker $w'$). In doing so, the controller uses the global barrier. Next, we describe how query-aware partitioning adapts over time.

**Dynamic Updates:** The global view about the query distribution is highly dynamic and workers have to keep the query information on the controller up-to-date. To this end, worker $w$ sends a $stats(q, |LS(q,w)|, I_w, w)$ message after each iteration to indicate that query $q$ has $|LS(q, w)|$ active vertices on worker $w$ and intersects with other queries according to the function $I_w : 2^\mathbb{Q} \to \mathbb{N}$. The intersection function returns the number of shared vertices between any combination of local query scopes. For example, if three queries $q_1, q_2$, and $q_3$ share three common vertices, the intersection function returns $I_w(\{q_1, q_2, q_3\}) = 3$. To increase communication efficiency, we piggyback statistics messages with barrier synchronization messages in our implementation.

The controller aggregates this information by calculating the global intersection functions of global query scopes based on the local intersections of local query scopes. It maintains all query statistics for a fixed duration, denoted as the (tumbling) *monitoring window*, given by the window parameter $\mu$ (cf. Section 2). The window parameter determines the degree of adaptivity of the system, i.e., how timely the partitioning reacts to dynamic query patterns. For instance, a larger window parameter allows older query statistics in the global view of the controller and leads to more long-term query-aware partitioning decisions (we specify the exact parameter choices for this and the following parameter in Section 4.1). Moreover, the controller initiates global barriers when the statistics indicate that the current partitioning is suboptimal. To this end, we use the *query locality*, i.e., the percentage of iterations which a query executes completely locally on a single worker, as a metric for the current quality of the partitioning. If the average query locality of all active queries is less than a threshold $\Phi$, the controller initiates a new partitioning. Note that the controller calculates the Q-cut algorithm (cf. Section 3.2) in parallel to the graph processing on the worker. Therefore, the partitioning latency is hidden and the duration of the global barrier is minimal (cf. Section 4).

## 4 EVALUATIONS

In this section, we show that adaptive Q-cut partitioning reduces query latency by up to 57% compared to the benchmarks. Moreover, we evaluate scalability and the hybrid barrier optimization.

### 4.1 Experimental Setup

For experiments validation, we implemented Q-Graph (25k lines of Java code) and made both source code and data publicly available.

**Graph Data and Query Generation:** For realistic graph data, we converted the OpenStreetMap road networks (i) *Germany* (GY)

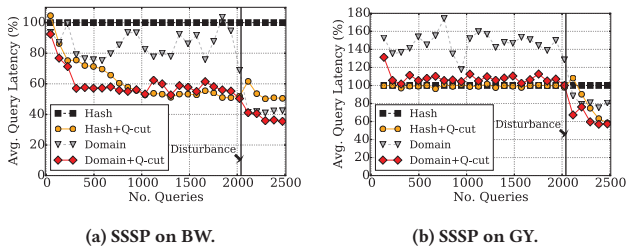

**Figure 5: (a)-(b) Adaptive query-aware partitioning reduces query latency over time.**

to a graph with 11,805,883 vertices (i.e., junctions) and 30,804,741 edges (i.e., road segments), and (ii) the region *Baden-Wuerttemberg (BW)* to a graph with 1,802,728 vertices and 4,770,566 edges. We set the edge weights to the length of the respective road segments, divided by the speed limit to estimate the travel time over this road segment. Although these are medium sized graphs, the goal in this paper is to reduce query latency in the presence of *multiple* localized and iterative queries. For instance, preliminary results on GraphX indicate more than three orders of magnitudes higher query latency compared to Q-Graph for the same problem instance.

We tested two queries: Single-source shortest path (SSSP) and Point-of-interest (POI). SSSP calculates the shortest path between a given start and end vertex. POI retrieves the closest vertex with a specified tag (e.g. gas station) to a given start vertex. We assign a tag to each vertex with probability $\frac{1}{12500}$ which is approximately the ratio of gas stations to road segments. To get realistic query workload, we determined the 64 biggest cities in GY and 16 biggest cities in BW and generated for each query a random start vertex around these *hotspots* – keeping the number of queries per city proportional to their populations. For SSSP, we also generated an end vertex with variable euclidean distance to the start vertex to account for intra- and inter-urban mapping queries.

**Initial Partitioning:** We used two algorithms with different strengths to initially partition the graph: *Hash* leads to ideal workload balancing, and *Domain* leads to ideal locality of up to 98% local execution per query (we validate both claims in Section 4.2). Domain serves as a best-case static partitioning algorithm: a domain expert, who already knows the hotspots of the query distribution in advance, manually partitions the graph such that each hotspot is assigned to a single partition. We also tested a state-of-the-art partitioning algorithm *linear deterministic greedy* (LDG) [36]. But LDG resulted in highly imbalanced partitions due to the skewness of the query distribution. Initial experiments with the imbalanced, LDG-partitioned graph suggest an increased average query latency by factor two to six compared to our methods. Hence, we excluded it from the experiments. Then, we applied Q-cut on top of this initial partitioning as queries are processed, and measured how query latency and performance changes over time.

**Computing Infrastructures:** To evaluate both scale-up and scale-out performance, we used the following computing infrastructures: (i) *M1*, a multi-core machine with 8 CPU cores (Intel(R) Core i7-2630QM, 2.90GHz, 6MB cache) and 8GB RAM, (ii) *M2*, a multi-core machine (AWS *m4.2xlarge*) with 8 CPU cores (Intel(R) Xeon E5-2676 v3, 2.4GHz, 30MB L3 cache) and 32GB RAM, and (iii) *C1*, a cluster with 8 nodes × 8 cores (Intel(R) Xeon(R), 3.0GHz, 6MB cache) and 32GB RAM per node, connected via 1-Gigabit Ethernet. For the scale-up infrastructures, we followed a well-established design choice to exploit $k$ cores by executing $k$ partitions in parallel on a single machine [41] and relied on loopback TCP for communication between partitions. To evaluate efficacy of query-cut partitioning, the scale-up infrastructure is a more challenging scenario than the scale-out infrastructure because the benefits of improved partitioning (i.e., reduced network traffic) are less pronounced on a single multi-core machine. Therefore, we first test QGraph in a scale-up (cf. Section 4.2) and then in a scale-out environment (cf. Section 4.3).

**System Settings:** QGraph has several system parameters that impact overall system performance. We list the most important ones in the following and published the full configuration to our open-source web repository (cf. Section 1) for reproducibility: (i) we set the time to calculate a query-cut on the controller to 2 seconds. (ii) If the controller detects that the average query locality is less than threshold $\Phi = 0.7$, it starts the Q-cut algorithm (cf. Section 3.4). Although we did not observe that query latency is sensitive to the exact parameter choice, we recommend a value between the locality of Hash and Domain (cf. Figure 6f), i.e., $\Phi \in [0.3, 0.99]$. For more global queries, a modest locality level of $\Phi < 0.5$ might be preferable. However, the parameter choice of $\Phi = 0.7$ is a robust decision for our localized queries in all performed evaluations. (iii) We set the monitoring window $\mu$ defining how long old queries will be considered (cf. Section 3.4) to $\mu = 240$ seconds in order to accumulate a few dozen queries in the Q-cut algorithm and restrict the maximal number of queries to 128. (iv) To reduce TCP overhead, the sender thread batches vertex messages with a maximum of 32 vertex messages per batch and 32 kilobytes batch size. Increasing the batch size beyond these limits has not reduced average query latency further due to the increasing waiting time on both the sender and receiver side.

### 4.2 Adaptive Q-cut Partitioning:

To show both static and dynamic behavior of the Q-cut algorithm, we executed 2048 hotspot SSSP queries on the BW graph in batches of 16 parallel queries ($k = 8$ workers on *M2*) – followed by a *disturbance* to test how Q-Graph adapts to changing query workloads (cf. Figure 5a). For the disturbance, we abruptly changed the query workload for further 496 queries from intra-urban to inter-urban queries between random neighboring cities. We measured average query latency and normalized by the query latency of Q-Graph using the static Hash partitioned graph. In the first phase of the experiment, Q-Graph with Q-cut reduces average query latency continuously by up to 49% compared to static Hash and by up to 40% compared to static Domain. The large fluctuation of Domain is a result of the increased imbalance of query workload compared to Hash and Q-cut.

In the second phase, the query workload changes: inter-urban queries become more complex with larger query scope. The low locality of static Hash harms scalability due to high communication overhead – the relative improvement of all other approaches compared to static Hash becomes more prominent. The Q-cut algorithm reduces average query latency on top of both static partitioning strategies Hash and Domain.

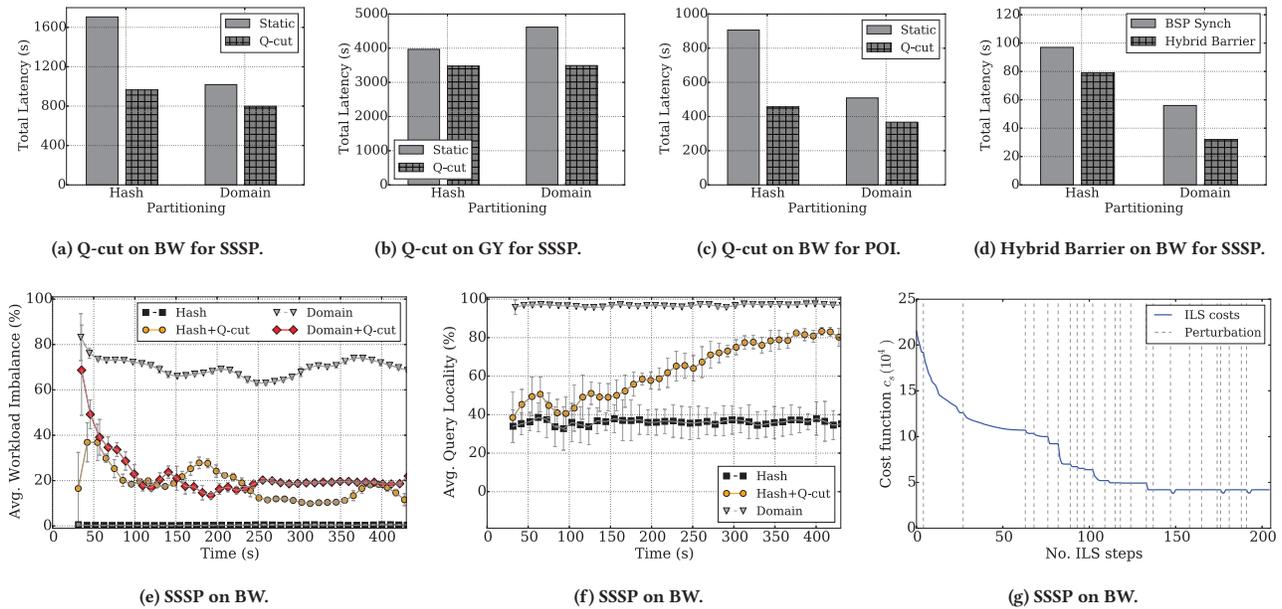

**Figure 6:** (a)-(b) Q-cut reduces query latency for different partitioning strategies for SSSP and (c) for POI compared to static partitioning Hash and Domain. (d) The query latency decreases with better partitioning and hybrid barrier synchronization. (e) Workload Balancing. (f) Percentage of local query executions. (g) Perturbation to overcome local minima.

We performed a similar experiment on *M2* using the GY graph in Figure 5b. In this setting, Q-cut reduces query latency by up to 45% compared to static Hash and 30% compared to static Domain partitioning. Note that for the larger GY graph, workload balancing is a more important objective than query locality as can be seen by the relatively improved performance of Hash. The reason is that the straggler problem becomes more prominent due to the higher complexity of the road network and higher number of queries processed by the worker responsible for the largest German city Berlin. However, Q-cut reduces average query latency again on top of both partitioning strategies Hash and Domain.

The summed latency over all 2048 queries is aggregated in Figure 6a for the SSSP query on the BW graph (reduced total latency by 43% compared to Hash and 22% to Domain), in Figure 6b for the SSSP query on the GY graph (reduced total latency by 13% compared to Hash and 25% to Domain). For the last experiment, we executed 2048 POI queries on the BW graph on *M2* to validate efficacy of Q-cut for different *types* of graph queries (cf. Figure 6c). The summed latency was reduced by 50% compared to static Hash and by 28% compared to static Domain.

Balancing the workload across machines mitigates the straggler problem and enables good resource utilization. We compare workload balancing of the four partitioning strategies in Figure 6e by executing 2048 SSSP queries on the BW graph (as previously described). We measured workload as the number of active vertices on a worker in a time window of 60 seconds and workload imbalance as a worker's deviation from the average workload (averaged over a sliding window of 10 seconds). Clearly, Domain leads to relative high workload imbalance because of the diverse query hotspots while Hash results in balanced workload. Q-cut converges to an imbalance of 20% because we set the maximally allowed imbalance to 25 percent, i.e., $\delta = 0.25$.

But how good is partitioning quality for the different partitioning algorithms? In Figure 6f, we measured the *query locality*, defined as the percentage of iterations that are executed completely locally. We calculated the running average and standard deviation over all queries with window size 20 seconds. The result shows that Domain leads to almost optimal locality of more than 95% while Hash reaches only 38% locality. However, even when Q-cut starts on top of the suboptimal Hash prepartitioning, locality increases over time and converges against a locality level of 80%. But, in contrast to Domain, Q-cut always ensures workload balance under dynamic query workload – higher query locality would result in higher workload imbalance which we do not allow.

Next, we validate the efficacy of the iterated local search algorithm in the controller. To this end, we plot the costs $c_s$ of the currently found best solution state *s* during a single run of the algorithm. We monitored the first execution of the ILS on the controller with the Hash-partitioned BW graph. The results show that costs $c_s$ are reduced by more than 75%. At the same time, the whole execution of the algorithm takes only two seconds – executed asynchronously to the graph query computation on the workers and hence introducing no latency penalty. The ILS execution retrieves a high-quality partitioning for a graph with millions of vertices in minimal time which is a consequence of our high-level query-centric representation. We also highlighted the points of perturbation (after ILS got stuck in local minima). The combination of local searches until convergence and perturbations provides an effective method to overcome local minima – which experimentally verifies the design decisions for the perturbation routine.

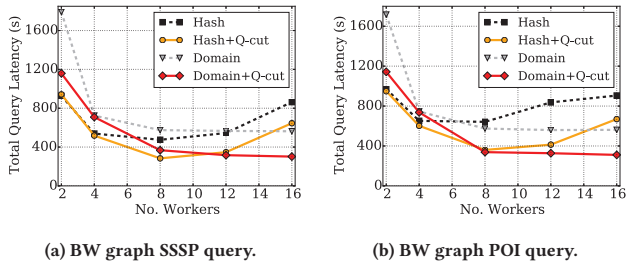

(a) BW graph SSSP query.  (b) BW graph POI query.

**Figure 7: Q-Graph scalability.**

### 4.3 Scalability and Hybrid Barrier

We evaluate how adding more workers impacts the total query latency of Q-Graph in Figure 7 for SSSP and POI on the BW graph by executing 1024 respective queries (16 parallel queries) on *C1*. For Hash prepartitioning on SSSP, the total query latency decreases from 927 to 474 seconds when increasing the number of workers from two to eight, and increases to 863 seconds when increasing the number of workers further. The reason is that Hash leads to high communication overhead in the distributed setting of *C1* and thus harms scalability for a large number of workers. This effect is alleviated by using Q-cut which reduces total query latency to 283 seconds for eight workers. For Domain prepartitioning on SSSP, the overall communication overhead is minimized which improves scalability – increasing the number of workers reduces latency from 1790 with two workers to 562 seconds with 16 workers. Q-cut leads to even better scalability reducing latency from 1150 with two workers to 301 seconds with 16 workers. The reason for Domain's high runtime for a lower number of workers (e.g. $k = 2$ workers) is the suboptimal workload balancing leading to massive straggler problems. Similar results were obtained for POI.

The hybrid barrier optimization enables full exploitation of locality. To show this, we measured total query latency of our Q-Graph system for 64 shortest path queries on the BW graph ($k = 8$ workers) on *M1* (cf. Figure 6d). We compared total query latency between traditional BSP-like barrier synchronization, i.e., all queries perform global synchronization after all iterations, and our hybrid barrier synchronization. Clearly, better partitioning (Domain) leads to $1.7 - 2.4\times$ reduced total query latency. But importantly, the hybrid barrier optimization reduces total query latency by $1.2 - 1.7\times$ for *both* methods: Domain and Hash.

### 5 RELATED WORK

Several graph processing systems have influenced and inspired our work. Pregel [22] was the first iterative graph system using the general-purpose, vertex-centric programming model on a message passing implementation. PowerGraph [12] suggests vertex-cut instead of edge-cut partitioning to improve performance on real-world graphs. PowerLyra [5] differentiates low- and high-degree vertices for processing and partitioning using a hybrid-cut algorithm. GraphX [43] is a library for graph-parallel computation on top of the Spark framework. GraphH [27] performs dynamic partitioning considering heterogeneous vertex traffic patterns. Mizan [19] proposes low-level vertex migration to balance workload dynamically using global synchronization barriers. We extended this idea to the hybrid barrier model that combines both local and global synchronization barriers. GPS [35] extends Pregel with decentralized dynamic repartitioning of vertices. All of the presented approaches support only a single, batch processing query rather than multiple, interactive graph queries. Hauck et al. [14] provided experimental evidence that single-query graph systems do not scale well in a multi-query environment.

Sedge [45] uses a complementary partitioning scheme based on replication of graph data to cope with localized queries and dynamic query hotspots. Parallelism is achieved by maintaining multiple independent Pregel instances. Hence, this is orthogonal to Q-cut, hybrid barriers, and adaptivity.

Several *query-agnostic* partitioning algorithms optimizing the number of cut vertices or edges have been proposed [7, 15, 23, 25, 30, 34, 36, 37, 42]. As shown in Section 4, even best-case edge-cut partitioning algorithms lead to suboptimal locality, workload balancing, and query latency.

Two graph systems support concurrent localized graph analytics queries. Weaver [9] focuses on concurrency control, i.e., how to enable transactional graph updates during query execution. Their interesting concept of refinable timestamps enables efficient ACID transactions on dynamic graphs. Seraph [44] decouples the data that is accessed by the individual queries and the computational logic such as the shared graph structure. However, Weaver and Seraph do not support adaptive partitioning which is the focus of our paper. Moreover, the existing body of research about graph database systems [1] does not specifically optimize query localization and synchronization in vertex-centric graph processing systems.

NScale [33] is a subgraph-centric graph processing system. Queries run in a $k$-hop neighborhood around a specified vertex. They consider replication to ensure that each subgraph is "completely contained within [...] one partition". In contrast, Q-Graph allows queries to dynamically grow and shrink on a shared graph structure. Hence, Q-Graph supports a superset of more complex graph applications.

### 6 CONCLUSION

Emerging CGA applications with localized graph queries running in parallel on a shared graph structure require novel types of partitioning and synchronization methods. We developed and evaluated the idea of scalable management of centralized knowledge about query workload to perform *query-aware adaptive partitioning*. Combined with a novel synchronization mechanism for CGA applications (*hybrid-barrier*), we observed a speedup of average query latency by up to 2.2×. Future work (i) examines query locality for algorithms such as localized PageRank on a billion-scale web graph with skewed degree distribution, (ii) explores query-based partial replication of vertices to reduce the query-cut size even more (cf. [28, 32]), and (iii) analyses whether distributing the centralized controller improves scalability even more.

## A APPENDIX: Q-CUT SUBROUTINES FOR ITERATED LOCAL SEARCH

For comprehensibility, we describe the four subroutines of the Q-cut algorithm, i.e., local search, perturbation, initial solution, and termination.

### A.1 Local Search Subroutine

We present our local search heuristic in Algorithm 2. The main idea is to simulate local query scope movements between workers as long as this leads to cost reductions. If no such move can be found, we return the last solution (denoted as *state*) which is a local minimum. More precisely, in lines 3-9, the algorithm iteratively determines the successor state $s'$ with minimal costs $c_{s'}$ and takes this state as a starting point for the next iteration until there is no successor state with smaller costs. In lines 13-17, the algorithm determines all possible successor states that would arise by moving any local query scope from worker $w$ to worker $w'$—for all respective combinations of workers and local query scopes. More precisely, as the number of these combinations can be very high, we *clustered* the queries as a preprocessing step into $4k$ clusters using a variant of the well-known Karger's algorithm with linear runtime complexity [16] and moved whole clusters between workers.

**Algorithm 2** Local search heuristic to find local minimum.

```
 1: function LOCALSEARCH(State s)
 2:     terminated ← False
 3:     while not terminated do
 4:         l ← SUCCESSORS(s)
 5:         s' ← argmin_{s'' ∈ l} c_{s''}
 6:         if c_{s'} < c_s then
 7:             s ← s'
 8:         else
 9:             terminated ← True
10:     return s
11: function SUCCESSORS(s)
12:     l ← ∅
13:     for w, w' ∈ W, q ∈ Q do
14:         x ← |LS(q, w)|
15:         if w ≠ w' ∧ x > 0 ∧ (|(L_w − x) − (L_{w'} + x)|)/(max(L_w − x, L_{w'} + x)) < δ then
16:             s' ← State after move(LS(q, w), w, w')
17:             l ← l ∪ {s'}
18:     return l
```

Balancing the workload is of major importance to efficiently utilize the system resources. We define workload as a combination of two metrics: the number of vertices assigned to worker $w$, i.e., $|V(w)|$, and the size of the local query scopes on worker $w$, i.e., $\sum_{q \in \mathbb{Q}} |LS(q, w)|$. We define workload $L_w$ as the combination of these scores, i.e., $L_w = \frac{|V(w)| + \sum_{q \in \mathbb{Q}} |LS(q, w)|}{2}$ and require that the workload of all pairs of workers is balanced, i.e., $\forall w, w' \in W : \frac{|L_w - L_{w'}|}{max(L_w, L_{w'})} < \delta$. We restrict the solution space to solutions with balanced workload by excluding successor states that would result in larger workload differences due to the movement of local query scopes in Algorithm 2 line 15.

### A.2 Perturbation Subroutine

An important subroutine of ILS is the method of perturbing a locally optimal solution state $s$. The perturbation transforms the converged state $s$ into a new state $s'$ that serves as a starting point for the next local search. A good perturbation is neither too small (i.e., the algorithm gets stuck in local minima), nor too large (i.e., the algorithm becomes uninformed). As a general hint, it was proposed to consider a meta representation of the solution and perform a random perturbation in this representation [21]. We used the following strategy with the idea of bringing together all local query scopes of a query to a single worker (cf. Figure 8).

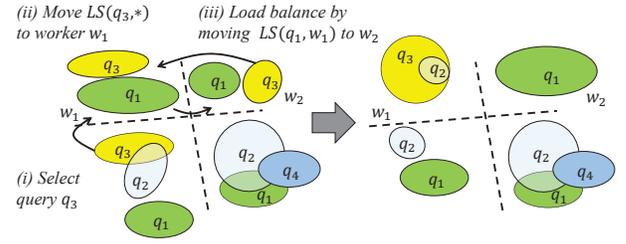

**Figure 8: Perturbation example.**

  I Randomly select query $q$ being spread across at least two workers.
 II Move local query scopes $LS(q, w'), w' \in W$ of query $q$ to the worker $w$ with the largest local query scope of query $q$.
III Balance workload by randomly moving local query scopes from the maximally to the least loaded worker until workload balancing is established.

This perturbation injects a certain amount of informed disorder (by merging local query scopes of a single query) but does not lead to a completely chaotic state (such as random restart).

### A.3 Initial Solution and Termination Subroutine

As initial solution, we take the current partitioning of the queries as received by the workers. For the termination criterion, we identified two requirements: (i) minimize the costs as much as possible to enable Q-Graph to fully exploit query locality, and (ii) prevent that the Q-cut algorithm becomes the bottleneck of the whole system. To this end, we define the termination criterion *outside of the ILS framework* by interrupting the computation as soon as a result is needed, i.e., when the adaptivity module decides to initiate repartitioning (cf. Section 3.4). The iterative nature of the Q-cut algorithm enables this early termination – even if the optimal solution is not yet reached. A major performance benefit comes from the decoupling of partitioning and computation logic: the controller can execute the Q-cut algorithm asynchronously at graph processing runtime using the latest *stats*-messages from the workers. Hence, if the controller initiates repartitioning, it can already propose a better Q-cut to the workers causing minimal latency overhead for Q-cut partitioning.